\documentclass{kluwer}    

\newdisplay{guess}{Conjecture}
\usepackage{graphicx}

\begin{document}  
\begin{article}
\begin{opening}         
\title{Eccentricities of Planets in Binary Systems}

\author{Genya \surname{Takeda}, Frederic A. \surname{Rasio}}

\runningauthor{Takeda \& Rasio}
\runningtitle{Eccentricities of Planets in Binary Systems}

\institute{
    Department of Physics and Astronomy, Northwestern University, 2145 Sheridan Road, Evanston, IL 60208, USA
    }
\date{September 30, 2005}

\begin{abstract}
The most puzzling property of the extrasolar planets discovered by  recent radial velocity surveys is their high orbital eccentricities,  which are very difficult to explain within our current theoretical  paradigm for planet formation. Current data reveal that at least 25\%  of these planets, including some with particularly high  eccentricities, are orbiting a component of a binary star system. The  presence of a distant companion can cause significant secular  perturbations in the orbit of a planet. At high relative  inclinations, large-amplitude, periodic eccentricity perturbations  can occur. These are known as ``Kozai cycles'' and their amplitude is  purely dependent on the relative orbital inclination. Assuming that  every planet host star also has a (possibly unseen, e.g., substellar)  distant companion, with reasonable distributions of orbital  parameters and masses, we determine the resulting eccentricity  distribution of planets and compare it to observations? We find that  perturbations from a binary companion always appear to produce an  excess of planets with both very high ($e\gsim 0.6$) and very low ($e \lsim 0.1$) eccentricities. The paucity of near-circular orbits in  the observed sample implies that at least one additional mechanism  must be increasing eccentricities.  On the other hand, the  overproduction of very high eccentricities observed in our models  could be combined with plausible circularization mechanisms (e.g.,  friction from residual gas) to create more planets with intermediate  eccentricities ($e\simeq 0.1-0.6$). 
\end{abstract}

\end{opening}

\section{Introduction}
As of September 2005, more than 160 extrasolar planets have been discovered by radial-velocity surveys\footnote{For an up-to-date catalog of extrasolar planets, see {\tt exoplanets.org} or {\tt www.obspm.fr/encycl/encycl.html}.}.
At least $\sim$10\% are orbiting a component of a wide stellar binary
system \cite{eggenberger04}. In contrast to the planets in our own solar system, one of the most remarkable properties of these extrasolar planets is their high orbital eccentricities.  These high orbital eccentricities are probably not significantly affected by observational selection effects \cite{fischer92}.  Thus, if we assume that planets initially have circular orbits when they are formed in a disk, there must be mechanisms that later increase the orbital eccentricity.  A variety of such mechanisms have been proposed \cite{tremaine04}.  Of particular importance is the Kozai mechanism, a secular interaction between a planet and a wide binary companion in a hierarchical triple system with high relative inclination \cite{kozai62,holman97,ford00}.  When the relative inclination angle $i_0$ between the orbital planes is greater than the critical angle $i_{\rm crit} = 39.2^\circ$ and the semimajor-axes ratio is sufficiently large (to be in a small-perturbation regime), long-term, cyclic angular momentum exchange occurs between the planet and the distant companion, and long-period oscillations of the eccentricity and relative inclination ensue.  To lowest order, the maximum of the eccentricity oscillation ($e_{1,\rm max}$) is given by a simple analytic expression: 
\begin{equation}
e_{\rm \max}\simeq\sqrt{1-(5/3)\cos^2{i_0}}
\end{equation}
\cite{innanen97,holman97}.
Note that $e_{\rm \max}$ depends just on $i_0$.  Other orbital parameters, such as masses and semimajor axes of the planet and the companion, affect only the period of the Kozai cycles.  Thus, a binary companion as small as a brown dwarf or even another Jupiter-size planet can in principle cause a significant eccentricity oscillation of the inner planet.  

Our motivation in this study is to investigate the possible global effects of the Kozai mechanism on extrasolar planets, and its potential to reproduce the unique distribution of observed eccentricities.  In practice, we run Monte Carlo simulations of hierarchical triple systems.  We have tested many different plausible models and broadly explored the parameter space of such triple systems.

\section{Methods and Assumptions}

The purpose of our study is to simulate the orbits of hierarchical triple systems and calculate the probability distribution of final eccentricities reached by the planet.  For each model, 5000 sample hierarchical triple systems are generated, with initial orbital parameters based on various empirically and theoretically motivated distributions, described below.  Our sample systems consist of a solar-type host star, a Jupiter-mass planet, and a distant~F-, G- or~K-type main-sequence dwarf (FGK dwarf) or brown dwarf companion.  The possibility of another giant planet being the distant companion is excluded since it would likely be nearly coplanar with the inner planet, leading to very small eccentricity perturbations.

The initial orbital parameters of the triple systems are randomly generated using the model distributions described in Table~\ref{table1}.  In this paper, we present six models, each with different initial conditions that are listed in Table~\ref{table2}.


\begin{table}
\begin{tabular}{lcc}
\hline
Parameter & Model Distribution Function & Ref. \\
\lcline{1-1}\rlcline{2-2}\rcline{3-3}

Host-star Mass....$m_0$ ($M_\odot$) & uniform in 0.9 - 1.3 $M_\odot$ & \\
Planet Mass........$m_1$ ($M_{\rm Jup}$) & uniform in $\log{m_1}, 0.3 - 10M_{\rm Jup}$ & [1] \\
Secondary Mass...$m_2$ ($M_\odot$) & $\xi(q\equiv m_2/m_1) \sim \ \exp\left\{{\frac{-(q-0.23)^2}{0.35}}\right\}$ & [2] \\
Semimajor Axis....$a_1$ (AU) & uniform in $\log{a_1}, 0.1 - 10\,$AU & [1], [3]  \\
of Planet & & \\
Binary Period.......$P_2$ (days) & $f(\log{P_2}) \sim \exp\left\{\frac{-(\log{P_2}-4.8)^2}{10.6}\right\}$ & [2] \\
Eccentricity of Planet...$e_1$ & $10^{-5}$ & \\
Age of the System...$\tau_0$ & uniform in $1-10\,$ Gyr & [4]\\
\hline
\end{tabular}
\caption{}[1] \inlinecite{zucker02}, [2] \inlinecite{duquennoy91}, [3] \inlinecite{ida04}, [4] \inlinecite{donahue98}\label{table1}
\end{table}


\begin{table}
\begin{tabular}{lrrrr}
\hline
\multicolumn{1}{r}{Model} & \multicolumn{1}{r}{$a_{2,\rm FGK}$} (AU) & \multicolumn{1}{r}{$a_{2,\rm BD}$$^a$} (AU) & \multicolumn{1}{r}{$e_2^b$} & \multicolumn{1}{r}{BDs$^c$} \\
\hline
A......... & using $P_2$, $<2000$ & $100-2000$ &$10^{-5}$ - 0.99 &
5\% \\
B......... & using $P_2$, $<2000$ & $100-2000$ & $10^{-5}$ - 0.99 &
10\% \\
C......... & using $P_2$, $<2000$ & $100-2000$ & $10^{-5}$ - 0.99 &
20\% \\
D......... & using $P_2$, $<2000$ & $100-2000$ & $10^{-5}$ - 0.99 &
30\% \\
E......... & ------------------ & $10-2000$ & 0.75 - 0.99 &
100\% \\
F......... & ------------------ & $10-2000$ & 0.75 - 0.99 &
5\% \\
\hline

\end{tabular}
\caption{}$^a$ uniform in logarithm \\
$^b$ all from thermal distribution, $P(e_2)=2e_2$ \\
$^c$ the fraction of brown dwarfs in 5000 samples 
\label{table2}
\end{table}


\begin{figure} 
\centerline{\includegraphics[width=16pc]{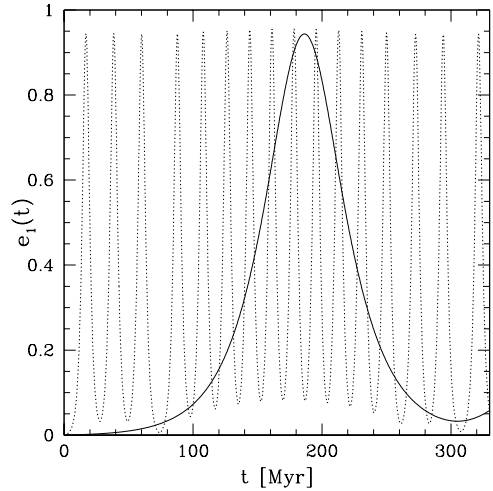}}
\caption{Eccentricity oscillation of a planet caused by a distant brown dwarf companion ($M=0.08M_\odot$, solid line) and by a main-sequence dwarf companion ($M=0.9M_\odot$, dotted line).  For both cases, the mass of the planet host star $m_0=1M_\odot$, the planet mass $m_1=1M_{\rm J}$, the planet semimajor axis $a_1=2.5\,$AU, the semimajor axis of the companion $a_2=750\,$AU, the initial eccentricity of the companion $e_2=0.8$, and the initial relative inclination $i_0=75^\circ$.  Note that $e_{1,\max}$ is the same in both cases, as it is dependent only on $i_0$, but the smaller mass of a brown dwarf companion results in a much longer oscillation period $P_{\rm KOZ}$.}
\label{twocycles}
\end{figure}

For the calculation of the eccentricity oscillations, we integrated the octupole-order secular perturbation equations (OSPE) derived in \inlinecite{ford00}.  These equations also include GR precession effects, which can suppress Kozai oscillations.  As noted by \inlinecite{holman97} and \inlinecite{ford00}, when the ratio of the Kozai period ($P_{\rm KOZ}$), to the GR precession period $(P_{\rm GR})$ exceeds unity, the Newtonian secular perturbations are suppressed, and the inner planet does not experience significant oscillation. 

Figure~\ref{twocycles} shows typical eccentricity oscillations in two different triple systems.  One contains a distant brown dwarf companion and the other a solar-mass stellar companion.  The two systems have the same initial orbital inclination $(i_0=75^\circ)$, and we see clearly that the amplitude of the eccentricity oscillation is about the same but with a much longer period $P_{\rm KOZ}$ for the lower mass companion.

To find the final orbital eccentricity distribution, each planetary orbit in our systems is integrated up to the assumed age of the system ($\tau_0$), and then the final eccentricity $(e_{\rm f})$ is recorded.  The results for representative models are compared to the observed eccentricity distribution in \S\ref{result}.  For more details, see \inlinecite{takeda05}.

\section{Results for the Eccentricity Distribution}\label{result}

\begin{figure} 
\centerline{\includegraphics[width=16pc]{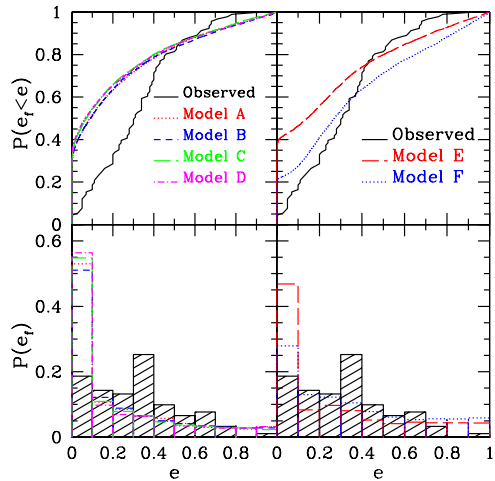}}
\caption{Final cumulative eccentricity distributions ({\it top}) and normalized probability distributions in histogram ({\it bottom}).  Four models with different fractions of brown dwarf and stellar companions.  Initial inclinations ($i_0$) are distributed uniformly in $\cos{i_0}$ ({\it left}).  Two extreme models where all the binary companions have orbits inclined by more than $40^{\circ}$ ({\it right}).}
\label{abcdef}
\end{figure}

Figure~\ref{abcdef} shows the final eccentricity in various models.  Each model is compared with the distribution derived from all the observed single planets with $a_1>0.1$, from the California \& Carnegie Planet Search Catalogue.  The statistics of the final eccentricities for our models and for the observed sample are presented in Table~\ref{stats}.
Models A-D represent planets in hierarchical triple systems with orbital parameters that are broadly compatible with current observational data and constraints on stellar and substellar binary companions.  All the models produce a large excess of planets with $e_{\rm f} < 0.1$ (more than 50\%), compared to only 19\% in the observed sample - excluding multiple-planet systems.  An excess of planets which remained in low-eccentricity orbits was evident in most of the models we tested.  Changing the binary parameters, such as separations or frequency of brown dwarf companions, did not change this result.


\begin{table*}
\label{stats}
\caption[]{Statistics of Eccentricity Distributions}
\begin{tabular}{lcccc}
\hline
Model & Mean & First Quartile & Median & Third Quartile \\
\hline
Observed ........ & 0.319 & 0.150 & 0.310 & 0.430 \\
A..................... & 0.213 & 0.000 & 0.087 & 0.348 \\
B..................... & 0.215 & 0.000 & 0.091 & 0.341 \\
C..................... & 0.201 & 0.000 & 0.070 & 0.322 \\
D..................... & 0.203 & 0.000 & 0.066 & 0.327 \\
E..................... & 0.245 & 0.000 & 0.141 & 0.416 \\
F..................... & 0.341 & 0.071 & 0.270 & 0.559 \\
\hline
\end{tabular}
\end{table*}


A major difference between most of the simulated and observed eccentricity distributions in the low-eccentricity regime ($e<0.1$) mainly arises from a large population of binary companions with low orbital inclination angle.  For an isotropic distribution of $i_0$, about 23\% of the systems have $i_0 < i_{\rm crit}$, leading to negligible eccentricity evolution. 
	For completeness, biased distributions of $i_0$ and $e_2$ are tested in model~E and F, as an attempt to achieve the best possible agreement with the observations.  With all the binary companions having sufficient inclination angles, model~F shows a better agreement with the observed sample in the low-eccentricity regime.  However, the number of planets remaining in nearly circular orbit ($e < 0.1$) is still larger than in the observed sample.  Moreover, model~F produces the largest excess of planets at very high eccentricities ($e>0.6$).  Note that these extreme models are clearly artificial, and our aim here is merely to quantify how large a bias would be needed to match the observations ``at any cost.''

\section{Summary and Discussion}

In most of our simulations, too many planets remain at very low orbital eccentricities.  The fraction of planets with $e<0.1$ is at least 25\% in our models, but only $\sim 15\%$ in the observed sample. There are several reasons for this overabundance of low eccentricities in our models.  First, the assumption of an isotropic distribution of $i_0$ automatically implies that $23\%$ of the systems have $i_0<i_{\rm crit}$, resulting in no Kozai oscillation.  This fraction already exceeds the observed fraction of nearly circular orbits ($e_1<0.1$) which is $\sim 15\%$.  Systems with sufficient initial relative inclination angles still need to overcome other hurdles to achieve highly eccentric orbits.  If many of the binary companions are substellar or in very wide orbits, Kozai periods become so long that the eccentricity oscillations are either suppressed by GR precession, or not completed within the age of the system (or both).  This can result in an additional 15\%-40\% of planets remaining in nearly circular orbits.  Even when the orbits of the planets do undergo eccentricity oscillations, there remains still 8-14\% that simply happen to be observed at low eccentricities.  Thus, our results suggest that the observed sample has a remarkably small population of planets in nearly circular orbits, and other dynamical processes must clearly be invoked to perturb their orbits. Among the most likely mechanisms is planet--planet scattering in multi-planet systems, which can easily perturb eccentricities to modest values in the intermediate range $\sim 0.2-0.6$ \cite{rasio96,weidenschilling96,marzari02}. Clear evidence that planet--planet scattering must have occurred in the $\upsilon$ Andromedae system has been presented by Ford, Lystad, \& Rasio (2005). Even in most of the systems in which only one giant planet has been detected so far, the second planet could have been ejected as a result of the scattering, or it could have been retained in a much wider, eccentric orbit, making it hard to detect by Doppler spectroscopy.

In the high-eccentricity region, where $e_1\gsim 0.6$, our models show much better agreement with the observed distribution.  The Kozai mechanism can even produce a small excess of systems at the highest eccentricities ($e_1>0.7$), although it should be noted that the observed eccentricity distribution in this range is not yet well constrained.  It is evident that the observed planets are rather abundant in intermediate values of eccentricity.  The Kozai mechanism tends to populate somewhat higher eccentricities, since during the eccentricity oscillation planets spend more time around $e_{1,\max}$ than at intermediate values.  However, this slight excess of highly eccentric orbits could easily be eliminated by invoking various circularization processes. For example, some residual gas may be present in the system, leading to circularization by gas drag \cite{adams03}.  In another scenario, decreased periastron distances can consequently remove the orbital energy of the planet by tidal dissipation.  This mechanism, referred to as ``Kozai migration'', was proposed by \inlinecite{wu03} to explain the orbit of HD80606 b.  
Kozai migration can also circularize the planetary orbit.  It is worth to note that the only three massive hot Jupiters in the observed sample, $\tau$Boo b, GJ86 b and HD195019 b ($M \sin{i} > 2 M_{\rm Jup}$, $P < 40$ days) are all in wide binary systems \cite{zucker02}.  Their tight orbits with low eccentricity can be a consequence of wider orbits with small periastron distances, initially invoked by the Kozai mechanism.

Clearly, even by stretching our assumptions, it is not possible to explain
the observed eccentricity distribution of extrasolar planets solely by
invoking the presence of binary companions, even if these companions are largely undetected or unconstrained by observations. However, our models suggest that
Kozai-type perturbations could play an important role in shaping the eccentricity distribution of extrasolar planets, especially at the high end. In addition,
they predict what the eccentricity distribution for planets observed around stars in wide binary systems should be.  The frequency of planets in binary systems is still very uncertain, but the search for new wide binaries among exoplanet host stars has been quite successful in the past few years (e.g., Mugrauer et al, 2005).

\acknowledgements
We thank Eric B.\ Ford for many useful discussions.
This work was supported by NSF grants AST-0206182 and AST-0507727.

\end{article}
\end{document}